\documentclass[aip,jcp,reprint]{revtex4-1}

\usepackage{graphicx}
\usepackage{amsmath}
\usepackage{latexsym}
\usepackage{hyperref}

\begin{document}

\noindent
Published as J. Chem. Phys. {\bf 141}, 094501 (2014).   

\url{http://dx.doi.org/10.1063/1.4894137}

\title{Packing frustration in dense confined fluids} 

\author{Kim~Nyg{\aa}rd}
\email[]{kim.nygard@chem.gu.se}
\affiliation{Department of Chemistry and Molecular Biology, University of Gothenburg, 
SE-412 96 Gothenburg, Sweden}

\author{Sten~Sarman}
\email[]{sarman@ownit.nu}
\affiliation{Department of Materials and Environmental Chemistry, 
Stockholm University, SE-106 91 Stockholm, Sweden}

\author{Roland~Kjellander}
\email[]{roland.kjellander@gu.se}
\affiliation{Department of Chemistry and Molecular Biology, University of Gothenburg, 
SE-412 96 Gothenburg, Sweden}

\date{\today}

\begin{abstract}
Packing frustration for confined fluids, i.e., the incompability between the preferred packing of the 
fluid particles and the packing constraints imposed by the confining surfaces, is studied for a dense 
hard-sphere fluid confined between planar hard surfaces at short separations. The detailed 
mechanism for the frustration is investigated via an analysis of the anisotropic pair distributions of 
the confined fluid, as obtained 
from integral equation theory for inhomogeneous fluids at pair correlation level 
within the anisotropic Percus-Yevick approximation. 
By examining the mean forces that arise 
from interparticle collisions around the periphery of each particle in the slit, we calculate the principal 
components of the mean force for the density profile  --  each component being the sum of collisional 
forces on a particle's hemisphere facing either surface. 
The variations of these components with the slit width give rise to rather intricate changes in the 
layer structure between the surfaces, but, as shown in this paper, the basis of these variations can 
be easily understood qualitatively and often also semi-quantitatively. It is found that the ordering of 
the fluid is in essence governed locally by the packing constraints at each single solid-fluid interface. 
A simple superposition of forces due to the presence of each surface gives surprisingly good 
estimates of the density profiles, but there remain nontrivial confinement effects that cannot 
be explained by superposition, most notably the magnitude of the excess adsorption of particles 
in the slit relative to bulk.

\end{abstract}

\maketitle

\section{Introduction}

Spatial confinement of condensed matter is known to induce a wealth of exotic crystalline 
structures.\cite{pieranski83,schmidt96,neser97,fortini06,fontecha08,oguz12}  
In essence this can be attributed to a phenomenon coined packing frustration; an incompability 
between the preferred packing of particles -- whether atoms, molecules, or colloidal particles -- 
and the packing constraints imposed by the confining surfaces. As an illustrative example, we 
can consider the extensively studied system of hard spheres confined between planar hard 
surfaces at a close separation of about five particle diameters or less. 
This is a convenient system for studies on packing frustration, because its phase 
diagram is determined by entropy only. While the phase diagram of the bulk hard-sphere system 
is very simple,\cite{pusey86} the dense packing of hard-sphere particles in narrow slits has been 
found to induce more than twenty novel thermodynamically stable crystalline phases, 
including exotic ones such as buckled and prism-like crystalline 
structures.\cite{schmidt96,neser97,fortini06,oguz12} 

In the case of spatially confined fluids, the effects of packing frustration are more elusive. 
Nevertheless, extensive studies on the fluid's equilibrium structure has brought into evidence 
this phenomenon; confinement-induced ordering of the fluid is suppressed when 
the short-range order preferred by the fluid's constituent particles is incompatible with the 
confining surface separation (see, e.g., Ref.~\onlinecite{nygard13} for illustrative examples).  
Packing frustration also influences other properties of the confined fluid, such as a strongly 
suppressed dynamics because of caging effects.\cite{mittal07,mittal08,lang10,lang12} 
However, little is known to date about the underlying mechanisms of frustration in spatially 
confined fluids. 

A stumbling block when elucidating the mechanisms of packing frustration in fluids is the hierarchy 
of distribution functions;\cite{hansen06} a mechanistic analysis of distribution functions 
requires higher-order distributions as input. While density profiles (i.e., singlet distributions) 
in inhomogeneous fluids are routinely determined today, either by theory, simulations, or 
experiments, structural studies are only seldom extended to the level of pair 
distributions.\cite{kjellander88a,kjellander91,kjellander91b,gotzelmann97,botan09,henderson97,zwanikken13} 
The overwhelming majority of theoretical work in the literature has been done on the singlet level where pair correlations from the homogeneous bulk fluid are used in various ways as approximations for the inhomogeneous system. Moreover, even in the cases where
the pair distributions for the inhomogeneous fluid have been explicitly determined,\cite{kjellander88a,kjellander91,kjellander91b,gotzelmann97,botan09} the 
mechanistic analysis of ordering is hampered by the sheer amount of variables. 
A conceptually simple scheme for addressing 
ordering mechanisms in inhomogeneous fluids is therefore much needed.  

In this work, we deal with the mechanisms of packing frustration in a dense hard-sphere 
fluid confined between planar hard surfaces by means of first-principles statistical mechanics at 
the pair distribution level. For this purpose we introduce principal components of the mean force 
acting on a particle, and study their behavior as a function of confining slit width. This provides a 
novel and conceptually simple scheme to analyze the mechanisms of ordering in inhomogeneous 
fluids. In contrast to the aforementioned multitude of exotic crystalline structures induced by packing 
frustration, we obtain compelling evidence that even for a dense hard-sphere fluid in narrow 
confinement, as studied here, the ordering is in essence governed by the packing constraints 
at a single solid-fluid interface. Nonetheless, there are also some common features for the 
structures in the fluid and in the solid phases. Finally, we demonstrate how subtleties in the 
ordering may lead to important, nontrivial confinement effects. 

The calculations in this work are done in integral equation theory for inhomogeneous fluids at pair correlation level, where the density profiles and anisotropic pair distributions are calculated self-consistently. The only approximation made is the closure relation used for the pair correlation function of the inhomogeneous fluid. We have adopted the Percus-Yevick closure, which is suitable for hard spheres. The resulting theory, called the Anisotropic Percus-Yevick (APY) approximation, has been shown to give accurate results for inhomogeneous hard sphere fluids in planar confinement.\cite{kjellander88b,kjellander91} In principle, pair distribution data for confined fluids could also be obtained from particle configurations obtained by computer simulation, e.g. Grand-Canonical Monte Carlo (GCMC) simulations. However, even with the computing power presently 
available, one would need impracticably long simulations in order to obtain a  reasonable 
statistical accuracy for the entire pair distribution, which for the present case has three independent variables. For the confined hard sphere fluid, the pair distribution function has narrow sharp peaks (see Ref.~\onlinecite{nygard13} for typical examples), which are particularly difficult to obtain accurately. The alternative to use, for example, the Widom insertion method to calculate the pair distribution point-wise by simulation is very inefficient for dense systems. It should be noted, however, that in cases where direct comparison is feasible in practice, simulations and anisotropic integral equation theory are in excellent agreement 
in terms of pair distributions.\cite{greberg97} In these cases, for a corresponding amount of pair-distribution data of essentially equal accuracy, the integral-equation approach was found to be many thousands times more efficient in CPU time than the simulations.
Finally we note that other highly accurate theoretical 
approaches, such as fundamental measure theory (see, e.g., Ref.~\onlinecite{roth10} for a 
recent review), have not yet been extended to the level of pair distributions in numerical applications.

\section{\label{sec:theory} System Description, Theory and Computations}

\begin{figure}
\centering\includegraphics[width=8cm]{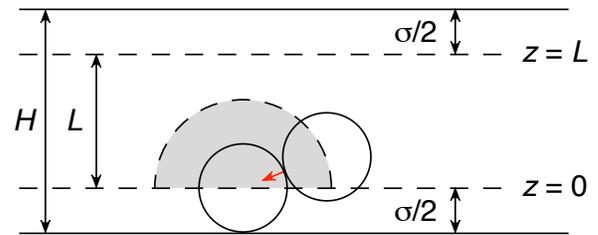}
\caption{Schematic of hard spheres between planar hard walls. The sphere diameter 
is denoted by $\sigma$, the surface separation by {\em H}, and the reduced slit width by {\em L}. 
The gray region depicts the excluded volume around the left particle, which in this figure is in 
contact with the bottom wall. The red arrow shows the collisional force exerted by the right particle 
on the left one. The force acts in the radial direction.
\label{fig:schematic}
} 
\end{figure}

Within the present study, we focus on a dense hard-sphere fluid confined between two 
planar hard surfaces. For a schematic representation of the confinement geometry, we refer to 
Fig.~\ref{fig:schematic}. The particle diameter is denoted by $\sigma$ and the surface separation 
by $H$. The space available for particle centers is given by the reduced slit width, which is defined 
as $L = H-\sigma$. The $z$ coordinate is perpendicular while the $x$ and $y$ coordinates are 
parallel to the confining surfaces. The system has planar symmetry and therefore the number 
density profile $n(z)$ only depends on the $z$ coordinate.

Except when explicitly stated otherwise, the confined fluid is kept in equilibrium with a bulk reservoir 
of number density $n_{\mathrm{b}} = 0.75\sigma^{-3}$. The average volume fraction of particles in 
the slit, $\phi_{\mathrm{av}} = (\pi \sigma^3/6H)\int_0^L n(z)dz$, then varies between about 0.33 
and 0.37 depending on the surface separation in the interval 
$L = 1.0\sigma-4.0\sigma$.\cite{nygard13}

Due to the planar symmetry all pair functions 
depend on three variables only, e.g., the pair distribution function 
$g({\bf r}_1,{\bf r}_2) = g(z_1,z_2,R_{12})$, where  
$R_{12}=|{\bf R}_{12}|$ with ${\bf R}_{12}=(x_2-x_1,y_2-y_1)$ denotes a distance parallel 
to the surfaces. In graphical representations of such functions, we let the $z$ axis go through 
the center of a particle at ${\bf r}_2$, i.e., we select  ${\bf r}_2 = (0,0,z_2)$. Then the 
function $g(z_1,z_2,R_{12})$ 
states the pair distribution function at position ${\bf r}_1 =({\bf R}_{12},z_1) = (x_1,y_1,z_1)$, given 
a particle at position $(0,0,z_2)$. Likewise, $n(z_1)g(z_1,z_2,R_{12})$ gives the average density 
at  ${\bf r}_1$ around a particle located at $(0,0,z_2)$.
We plot for clarity also negative values of $R_{12}$, i.e., in the following plots of pair functions 
$R_{12}$ is to be interpreted as a coordinate along a straight line in the $xy$ plane through the 
origin.

Throughout this study, we make use of integral equation theory for inhomogeneous fluids on 
the anisotropic pair correlation level. 
Following Refs.~\onlinecite{kjellander91} and \onlinecite{kjellander88b}, we determine the 
density profiles $n(z_1)$ 
and pair distribution functions $g(z_1,z_2,R_{12})$ of the confined hard-sphere fluid by solving 
two exact integral equations self-consistently: the 
Lovett-Mou-Buff-Wertheim equation, 
\begin{equation}
\frac{d [{\mathrm{ln}} n(z_1)+\beta v(z_1)]}{dz_1}=
 \int c(z_1,z_2,R_{12})\frac{dn(z_2)}{dz_2} dz_2 d{\bf R}_{12},  
\label{eq:LMBW}
\end{equation} 
and the inhomogeneous Ornstein-Zernike equation,  
\begin{equation}
h({\bf r}_1,{\bf r}_2)=c({\bf r}_1,{\bf r}_2)+
\int h({\bf r}_1,{\bf r}_3)n(z_3)c({\bf r}_3,{\bf r}_2) d{\bf r}_3, 
\label{eq:OZ}
\end{equation}
where $h=g-1$ is the total and $c$ the direct pair correlation function, while $v$ denotes the hard 
particle-wall potential 
\begin{equation} 
v(z) = \left\{ \begin{array}{ll} 
0 & \textrm{if $0 \leq  z  \leq L$,}\\ 
\infty & \textrm{otherwise.}
\end{array} \right. 
\label{eq:particle_wall}
\end{equation} 
As the sole approximation, we thereby make use of the Percus-Yevick  closure for anisotropic 
pair correlations, $c = g - y$, where $y({\bf r}_1,{\bf r}_2)$ denotes the cavity correlation function that 
satisfies $g = y \exp (-\beta u)$ and $u$ is the hard particle-particle interaction potential, 
\begin{equation} 
u({\bf r}_1,{\bf r}_2) = \left\{ \begin{array}{ll} 
0 & \textrm{if $\vert {\bf r}_1 - {\bf r}_2 \vert \geq \sigma$,}\\ 
\infty & \textrm{if $ \vert {\bf r}_1 - {\bf r}_2 \vert < \sigma$.}
\end{array} \right.
\label{eq:particle}
\end{equation} 
This set of equations constitutes the APY theory.

The confined fluid is kept in equilibrium with a bulk fluid reservoir of a given density by means of  
a special integration routine, in which the rate of change of the density profile for 
varying surface separation is given by 
the exact relation\cite{kjellander91,kjellander88b}
\begin{multline}
\frac{\partial n(z_1;L)}{\partial L}=-\beta n(z_1;L) \bigg[  \frac{\partial v(z_1;L)}{\partial L} \\
+ \int n(z_2;L)h(z_1,z_2,R_{12};L)\frac{\partial v(z_2;L)}{\partial L} dz_2 d{\bf R}_{12} \bigg]  
\label{eq:profile_equil}
\end{multline}
under the condition of constant chemical potential.
Here we have explicitly shown the $L$ dependence of the functions, which is implicit in the 
previous equations. For a concise review of the theory and details on the computations we refer 
to Ref.~\onlinecite{nygard13}.

\begin{figure}
\centering\includegraphics[width=8.5cm]{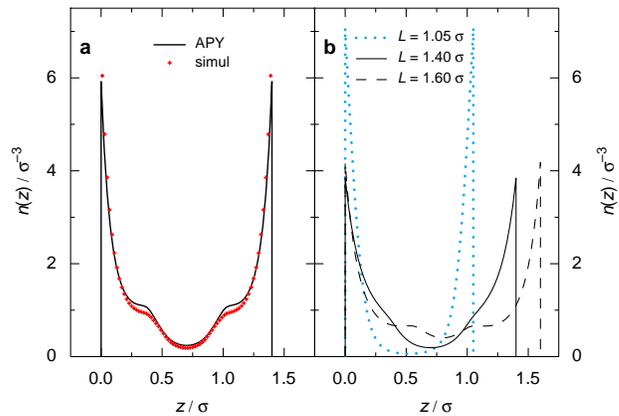}
\caption{Number density profiles $n(z)$ for the hard-sphere fluid confined between hard 
planar surfaces. 
(a) Data for the average volume fraction $\phi_{\mathrm{av}} = 0.40$ of particles in the slit of width 
$H=2.4\sigma$ (reduced slit width $L=1.4 \sigma$), which is virtually at phase separation to the 
crystalline state for this surface separation. 
The solid line depicts theoretical data within the Anisotropic Percus-Yevick (APY) approximation, 
while the crosses show data from the Grand-Canonical Monte Carlo simulation 
of Ref.~\onlinecite{mittal07}.     
(b) Theoretical data from APY approximation for a confined fluid in equilibrium with a bulk 
reservoir of number density $n_{\mathrm{b}} = 0.75\sigma^{-3}$. The reduced slit widths are 
$L = 1.05\sigma$ (dotted line), $1.40\sigma$ (solid line), and $1.60\sigma$ (dashed line). The 
average volume fraction $\phi_{\mathrm{av}}$ is here 0.35, 0.34, and 0.33, respectively.
\label{fig:simul}
} 
\end{figure}

\section{\label{sec:results}Results and Discussion}

\subsection{Density profiles and pair densities}

The theoretical approach adopted in this work has recently been shown\cite{nygard12} to be 
in quantitative agreement with experiments at the pair distribution level for a confined hard-sphere 
fluid in contact with a bulk fluid of the same density, $n_{\mathrm{b}} = 0.75\sigma^{-3}$, as used 
in the current work. Both the anisotropic structure factors from pair correlations and the density 
profiles agree very well with the experimental data. For higher densities, we compare in 
Fig.~\ref{fig:simul}(a) our result with the density profile obtained from GCMC  simulations by 
Mittal et al.\cite{mittal07} for an average volume fraction in the slit 
$\phi_{\mathrm{av}} = 0.40$ and at $L=1.40\sigma$. For this extreme particle density, which is 
virtually at phase separation to the crystalline phase at this surface separation,\cite{fortini06} there 
are quantitative differences, but our theoretical profile agree semi-quantitatively with the simulation 
data. For $L=1.0\sigma$ and $2.0\sigma$ and at the same  $\phi_{\mathrm{av}}$, the deviations 
between our profiles and the GCMC profiles by Mittal et al. are larger (not shown). In the rest of 
this paper we shall, however, treat cases with lower particle concentrations in the slit: 
$\phi_{\mathrm{av}}$ between about 0.33 and 0.37, which are less demanding theoretically. 
In Ref.~\onlinecite{kjellander88b} we showed, for a wide range of slit widths, that our theory is 
in very good agreement with GCMC simulations for the confined hard sphere fluid in equilibrium 
with a bulk density $0.68\sigma^{-3}$, which is only slightly lower than what we consider in this 
work. Furthermore, our main concern in this paper are cases with surface separations about 
halfway between integer multiples of sphere diameters, as in Fig.~\ref{fig:simul}(a).

Returning to the system in equilibrium with a bulk with density $n_{\mathrm{b}} = 0.75\sigma^{-3}$, 
we illustrate the concept of packing frustration in spatially confined fluids by presenting the number 
density profile $n(z)$ for reduced slit widths of $L=1.05\sigma$, $1.40\sigma$, and $1.60\sigma$ 
in Fig.~\ref{fig:simul}(b). The fluid in the narrowest slit exhibits strong ordering, as illustrated by 
well-developed particle layers close to each solid surface. Such ordering is observed for the 
hard-sphere fluid in narrow hard slits when the surface separation is close to an integer multiple 
of the particle diameter $\sigma$. 
In this specific case, the average volume fraction $\phi_{\mathrm{av}} = 0.35$ is about $82\%$  of the volume fraction for phase separation to the crystalline phase at this surface separation,\cite{fortini06} 
and the ``areal'' number density near each solid surface is 
$\int_0^{L/2} n(z)dz \approx 0.69\sigma^{-2}$, about $77\%$ of the freezing density for the 
two-dimensional hard-sphere fluid.\cite{schmidt96,fortini06}

For slit widths intermediate between integer multiples of $\sigma$,  the confined fluid develops into 
a relatively disordered fluid in the slit center, despite the confining slit being narrow enough to 
support ordering across the slit. In the particular case shown in Fig.~\ref{fig:simul}(b), we 
observe shoulders in the density peaks close to each surface which evolve with increasing slit 
width into two small (or secondary) density peaks in the slit center. At slightly larger slit widths 
(to be investigated below), these two small peaks merge and form a fairly broad layer in the middle 
of the slit. For $L \approx 2.0\sigma$ there is strong ordering again; the layers at either wall are 
then very sharp and the mid-layer is quite sharp (more profiles for $L=1.0\sigma-4.0\sigma$ for 
the current case can be found in Ref.~\onlinecite{nygard13} and as a video in 
Ref.~\onlinecite{note_film}). 
Such a change of ordering at 
the intermediate separations is usually interpreted as a signature of packing frustration, and in 
this paper we will address its mechanisms.

\begin{figure*}
\centering\includegraphics[width=14cm]{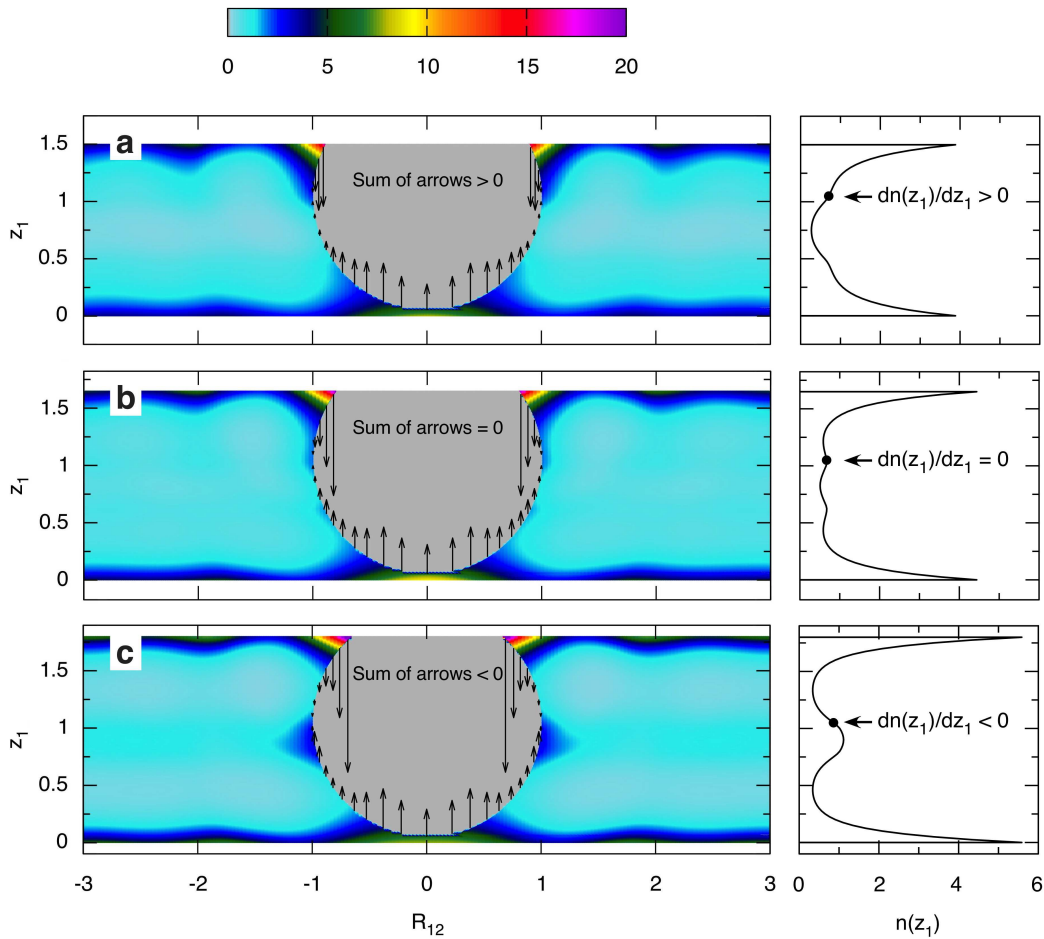}
\caption{Contour plot of the pair density 
$n({\bf r }_1)g({\bf r}_1,{\bf r}_2) \equiv n(z_1)g(z_1,z_2,R_{12})$ at coordinate 
${\bf r}_1=({\bf R}_{12},z_1)$ around a particle in the slit between two hard surfaces, when 
the particle is located on the $z$ axis at coordinate ${\bf r}_2  =(0,0,z_2)$. One surface is 
$0.5\sigma$ above the top and one $0.5\sigma$ below the bottom of each subplot (cf. 
Fig.~\ref{fig:schematic}). The system is in equilibrium with a bulk fluid of density 
$n_{\mathrm{b}} = 0.75\sigma^{-3}$ [same as in Fig.~\ref{fig:simul}(b)]. 
The gray region is the excluded volume zone around the particle. 
Data are shown for different reduced slit widths: (a) $L = 1.50\sigma$, (b) $1.65\sigma$, 
and (c) $1.80\sigma$. The number density profile $n(z_1)$ for each case is also shown for 
clarity to the right. 
The particle position $z_2$ (shown as filled circle in the profile plots) is in all cases positioned 
at a distance of $1.55\sigma$ from the bottom surface (at $z$ coordinate $1.05\sigma$). The 
arrows in the gray region depict  $z$ components of the collisional forces acting on the 
particle (corresponding to the $z$ projection of the red arrow in Fig.  \ref{fig:schematic}). The 
arrows displayed at a certain $z_1$ coordinate here represent the entire force acting on the 
sphere periphery in a $dz$ interval  around this coordinate. In subplot (a) the sum of all arrows 
(with signs) is $>0$, in (b) $=0$ and in (c) $<0$.
\label{fig:ng}
} 
\end{figure*}

How can we understand these observations? The starting point for our discussion will be the 
pair density $n({\bf r }_1)g({\bf r}_1,{\bf r}_2)$, i.e., the density at position ${\bf r}_1$ given a 
particle at position ${\bf r}_2$.  As will become evident below, the pair densities allow us to 
analyze the mechanisms leading to the detailed structure of the layers in confined, 
inhomogeneous fluids. Here, we shall in particular investigate the mechanisms of packing 
frustration in dense hard-sphere fluids under spatial confinement. 

Fig.~\ref{fig:ng} shows examples of contour plots of the pair density $n(z_1)g(z_1,z_2,R_{12})$ 
for three reduced slit widths, $L = 1.50\sigma$, $1.65\sigma$, and $1.80\sigma$, when a particle 
(the ``central'' particle) is located on the $z$ axis at coordinate ${\bf r}_2  =(0,0,z_2)$. The density 
profiles for these three slit widths are shown in the right hand side of the plot. In Fig.~\ref{fig:ng}(a)  
there is a shoulder in the profile on either side of the midplane, while in Fig.~\ref{fig:ng}(b) two small, 
but distinct, peaks have formed near the slit center. In Fig.~\ref{fig:ng}(c) these 
secondary peaks 
have merged into one peak in the middle. These changes in the density profile occur within a 
variation in surface separation of only $0.3\sigma$. In the contour plots, the central particle's 
center is in all cases situated at a distance of $1.55\sigma$ (about three particle radii) from the 
bottom surface, $z_2 = 1.05\sigma$, marked by a filled circle in the profile. Particles that form the main 
layer in contact with the bottom surface can then touch the central particle; the latter is penetrating 
just the edge of this layer. Note that the position of the secondary maximum for the middle case, 
Fig.~\ref{fig:ng}(b), is also located at $z_1 = 1.05\sigma$.

Thus, it can be understood that particles forming the small secondary peak in Fig.~\ref{fig:ng}(b) 
are in contact with, but barely penetrating, the main layer of particles at the bottom surface. The 
particles of this secondary peak are at the same time strongly penetrating the main layer at the 
top surface. The same is, however, true for the particles around the same $z$ coordinate 
($1.05\sigma$) in Figs.~\ref{fig:ng}(a) and \ref{fig:ng}(c), but with a markedly different outcome 
for the profile. Our task here is to understand the reason for such differences. 

In the contour plots of the pair density $n(z_1)g(z_1,z_2,R_{12})$ in Fig.~\ref{fig:ng} we see 
that in all three cases the particle density 
in the wedge-like section formed between the central particle and the upper wall is strongly enhanced, 
resulting in a local number density of up to 17, 20, and 24 $\sigma^{-3}$, respectively, in the 
three cases. This enhancement in $ng$ relative to the density $n$ at the same $z$ coordinate 
is given by the pair 
distribution function $g$, which is about 4 -- 4.5 in the inner part of the wedge-like 
section for all these cases. In the region near the bottom surface, where the central particle is in contact 
with the main bottom layer, there is also an enhancement in density, but to much smaller extent 
than at the top. Note that the density distribution $ng$ near the bottom is very similar in all three 
cases.

\subsection{Mean force}

To understand why the profiles differ so much in these three cases, we investigate the mean force 
$F(z)$ that acts on a particle with its center at position $z$. The potential of mean force, $w$, is 
related to the density $n$ by $n=n_{\mathrm{b}}\exp(-\beta w)$, where 
$\beta = (k_B T)^{-1}$, $k_B$ is Boltzmann's constant, and 
$T$ the absolute temperature. This implies that $F \equiv -\nabla w = k_B T \nabla \ln n$. Due to
 the planar symmetry, $n$ depends on $z$ only and the total force components in the $x$ and $y$ 
directions are zero. The mean force $F$ is then directed parallel to the $z$ axis and we have 
$\beta F(z) = d \ln n(z)/dz = n'(z)/n(z)$, where $n'=dn/dz$. Thus, an understanding of the 
behavior of the profile can be obtained from an analysis of $F$. The sign of $F$ tells whether
$n$ is increasing or decreasing and extremal points of $n$ correspond to points where $F$ is zero. 

For hard-sphere fluids, the forces exerted on a particle by the other particles in the system are 
simply due to collisions. Due to planar symmetry, the density distribution in the vicinity of a 
particle has rotational symmetry around the $z$ axis through the particle center. The average 
force at each point is acting in the direction normal to the 
sphere surface and for a particle located at $z_2$ the average force from all collisions 
along the sphere periphery at coordinate $z_1$ is 
proportional to the contact density $n(z_1)g_{\mathrm{cont}}(z_1,z_2)$, where 
$g_{\mathrm{cont}}(z_1,z_2)  \equiv g(z_1,z_2,R_{12})|_{R_{12}^2 = \sigma^2 - (z_1-z_2)^2}$ 
is the contact value of the pair distribution function at the particle surface. Note that the force in, for 
example, the $x$ direction on one side of the periphery is cancelled by the force in the $-x$ direction 
on the opposite side. Thus only the $z$ component of the net force on the particle contributes as 
expected. Using the first Born-Green-Yvon equation one can show~\cite{kjellander91} that 
\begin{equation}
\beta F(z_2)= 2\pi \sigma \int^{z_2+\sigma}_{z_2-\sigma}  n(z_1)g_{\mathrm{cont}}(z_1,z_2)
\frac{(z_2-z_1)}{\sigma} dz_1
\label{eq:BGY} 
\end{equation}
(the integral is over the range $|z_2-z_1| \leq \sigma$ where $g_{\mathrm{cont}}$ is 
defined).  The role of the factor $(z_2-z_1)/\sigma$ is to project out the $z$ component of the 
contact force. (This line of reasoning is readily extended to systems exhibiting soft interaction 
potentials, such as Lennard-Jones fluids or electrolytes; in such cases, however, one also needs 
to include the interactions with the walls and all other particles in the system, see e.g. 
Refs. \onlinecite{kjellander91b} and \onlinecite{kjellander09}.) 

Let us now return to the intriguing formation of secondary density maxima for $L \approx 1.65\sigma$. 
For this purpose, we present in Fig.~\ref{fig:ng} the $z$ component of the contact forces acting 
on the particle. They are represented by the arrows along the sphere periphery. 
In these plots, there are two major contributions to the net force acting on the particle, namely 
the repulsive forces exerted by the particle layers close to each confining wall. For  $L=1.65\sigma$ 
and  the chosen position of the central particle in Fig.~\ref{fig:ng}(b), $z_2 = 1.05\sigma$, these 
force contributions cancel each 
other: the sum of the arrows (with signs) is zero and hence $dn/dz=0$ at this $z$ coordinate, as 
shown to the right in the figure. It is the subtle interplay between these forces for neighboring 
$z_2$ values 
which leads to the secondary density maximum. 

The situation is, however, markedly different for 
$L=1.50\sigma$ and $1.80\sigma$. While the total force exerted by the particles in the main layer 
at the bottom surface is practically equal for all three cases, the magnitude 
of the force exerted by the particles in the main layer at the upper surface varies strongly with $L$.
This variation is partly due to the different magnitude of the contact densities in the wedge-like 
region mentioned above and partly due to the change in angle between the normal vector 
to the sphere surface there and the $z$ axis. Recall that the contact force acts 
along this normal vector, so the $z$ component is dependent on this angle.
For $L=1.50\sigma$, Fig.~\ref{fig:ng}(a), the  $z$ component of the contact force from the upper 
layer is smaller than for $L=1.65\sigma$. The sum of the arrows is then positive, i.e. the total 
average force is directed towards the upper wall and hence $dn/dz>0$ at this $z$ coordinate. 
For $L=1.80\sigma$, Fig.~\ref{fig:ng}(c), this $z$ component is larger compared to 
$L=1.65\sigma$, thereby pushing the central particle towards the slit center. Hence $dn/dz<0$ 
at this $z$ coordinate.

\begin{figure}
\centering\includegraphics[width=8.0cm]{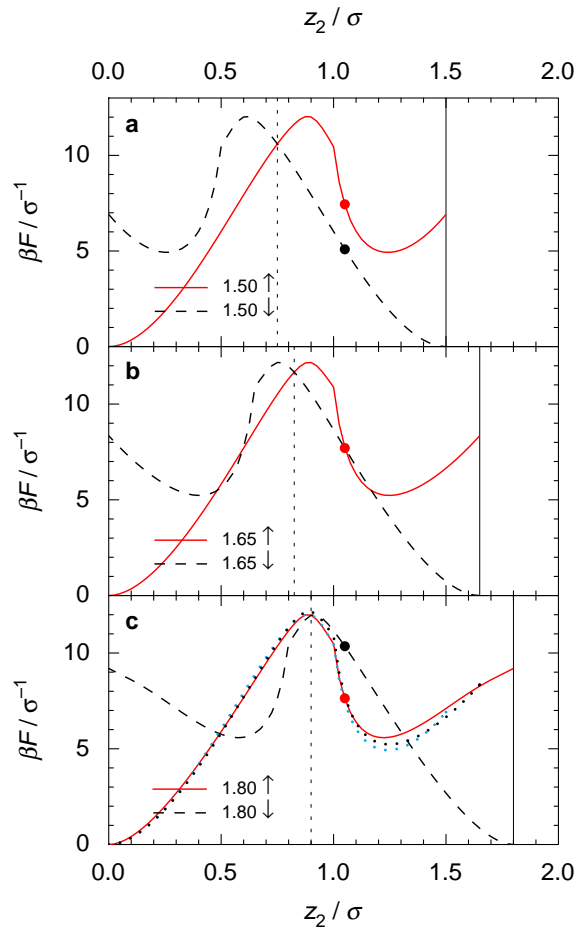}
\caption{Net forces acting on a particle for the systems in Fig.~\ref{fig:ng}. The force 
contributions acting in positive 
($F_{\uparrow}$, full curve) and negative ($F_{\downarrow}$, dashed curve) $z$ directions 
are presented separately as 
functions of particle position $z_2$ across the confining slit. Data are shown for reduced slit 
widths $L=1.50\sigma$, $1.65\sigma$, and $1.80\sigma$. The values of the forces for a particle 
at the $z_2$ coordinates in Fig.~\ref{fig:ng} are shown by filled circles. The dashed vertical line 
denotes the slit center, while the solid vertical line on the right-hand side indicates the upper 
limit for possible $z_2$ coordinates of the particle in the slit. 
For comparison of all three cases, $F_{\uparrow}$ is also shown for $L=1.50\sigma$ (blue dots) 
and $1.65\sigma$ 
(black dots) in the bottom panel. 
\label{fig:force}
} 
\end{figure}

\subsection{Principal components of mean force\label{sec:Principal_comp}}

In order to gain more insight into the formation of the secondary maxima, we present in 
Fig.~\ref{fig:force} the net force acting on a particle for all positions $z_2$ in the same three 
cases as discussed above, $L=1.50\sigma$, $1.65\sigma$, and $1.80\sigma$. To facilitate the 
interpretation, the principal force contributions acting in positive (denoted $F_{\uparrow}$) 
and negative ($F_{\downarrow}$) directions are shown separately. The total net force is 
$F = F_{\uparrow}-F_{\downarrow}$, where $F_{\uparrow}$ originates from collisions on 
the lower half of the sphere surface and $F_{\downarrow}$ on the upper half
($F_{\uparrow}$ and $F_{\downarrow}$ correspond to the absolute values of the sums 
of arrows in respective hemisphere in Fig.~\ref{fig:ng}). In Fig.~\ref{fig:force} 
the red (solid) and black (dashed) curves are each other's mirror images with respect 
to the vertical dashed axis at $z_2=L/2$, which shows the location of the slit center.  

Since the variation in $F_{\uparrow}$ (and $F_{\downarrow}$) is very similar for all slit widths in 
Fig.~\ref{fig:force}, the following discussion will hold for all three cases. For $z_2=0$ we have 
$F_{\uparrow} = 0$, because no spheres can collide from below since the confining surface 
precludes them from being there (cf. Fig.~\ref{fig:schematic}). With increasing $z_2$ we observe 
a monotonically increasing 
$F_{\uparrow}$, which can be attributed both to the increasing area exposed to collisions on 
the lower half of the sphere surface and the decrease in angle of the sphere normal there relative 
to the $z$ axis. With further increase in $z_2$, we eventually observe a decrease in the exerted 
force induced by a decrease in contact density $n(z_1)g_{\mathrm{cont}}(z_1,z_2)$. 
Around $z_2 \approx 1.0\sigma$ we observe a sudden onset of a rapid decrease for  
$F_{\uparrow}$. This is a consequence of a rapid decrease in contact density, that occurs when 
the particle at $z_2$ loses contact with the dense particle layer at the bottom wall. 
For even larger $z_2$, where the particle is close to the top surface, collisions with particles 
in the slit center around the entire lower half of the sphere surface become important so that 
$F_{\uparrow}$ increases again. 

The three red curves are compared in the bottom panel, where $F_{\uparrow}$ from the first 
two panels ($L=1.50\sigma$ and $1.65\sigma$) are shown as dotted curves. We see 
that the curves are nearly identical apart from in a small region to the far right. 
The analogous statement is true, of course, for the black dashed curves. 
Thus, apart from small $z_2$ intervals to the extreme left and right, the behavior of 
$F = F_{\uparrow}-F_{\downarrow}$ for the different slit widths can be understood in terms 
of a horizontal 
shift of the red and the black curves relative to each other. The formation of secondary density 
maxima can then 
be explained from the resulting balance of the force contributions. For $L = 1.65\sigma$ and 
$z_2>L/2$ [i.e., the right half of Fig.~\ref{fig:force}(b)], the two curves intersect at two points where 
the forces cancel each other and where $dn/dz=0$. The 
intersection marked by the filled circle gives a local maximum of $n(z)$ and the next one 
to the right gives a minimum. Together with the minimum at the slit center, $z_2=L/2$, where the 
curves also cross each other, these features give rise to the 
secondary peak of the density profile as we have seen in Fig.~\ref{fig:ng}(b). This subtle balance 
of forces, and hence the formation of secondary maxima, is only observed in a narrow range of 
slit widths, as evidenced by the force profiles for $L=1.50\sigma$ and $1.80\sigma$. In the latter 
case, the intersection at $z_2=L/2$ corresponds to a local maximum and the other one to a 
minimum. Together they give one peak in the middle as seen in Fig.~\ref{fig:ng}(c). 

The formation of secondary maxima is for $z_2>L/2$ accordingly a consequence of two phenomena: 
(i) the rapid decrease of $F_{\uparrow}$ followed by the subsequent increase of $F_{\uparrow}$
and (ii) the monotonic decrease of $F_{\downarrow}$ in the same region. Together these effects 
lead to the force curves intersecting twice in the manner they do for $L = 1.65\sigma$. The rapid 
decrease of $F_{\uparrow}$ is, as we have seen, due to the loss of contact of the particle with 
the well-developed bottom layer, while the monotonic decrease of $F_{\downarrow}$ occurs when 
the particle approaches the top surface.

\begin{figure}
\centering\includegraphics[width=8.0cm]{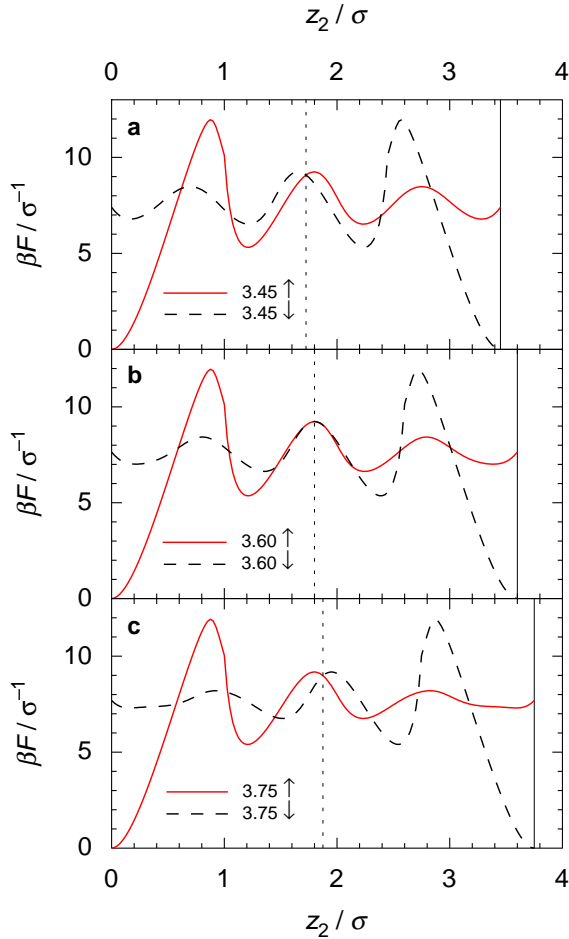}
\caption{As Fig.~\ref{fig:force}, but for reduced slit widths $L=3.45\sigma$, $3.60\sigma$, 
and $3.75\sigma$. 
\label{fig:force2}
} 
\end{figure}

\begin{figure}
\centering\includegraphics[width=8.5cm]{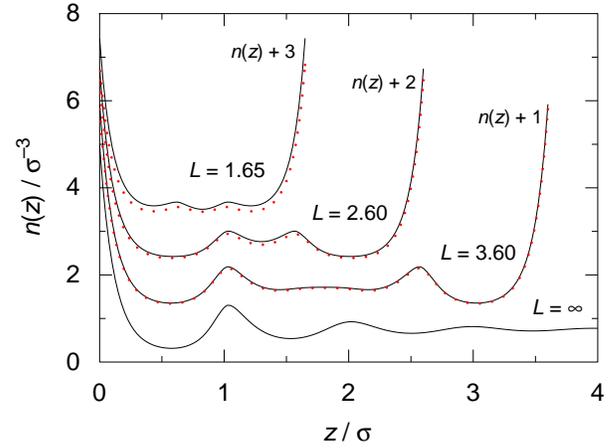}
\caption{Number density profiles $n(z)$ for the confined hard-sphere fluid. 
The reduced slit widths are $L = 1.65\sigma$ (offset vertically by 3.0$\sigma^{-3}$), 
$2.60\sigma$ (offset by 2.0$\sigma^{-3}$), and $3.60\sigma$ (offset by 1.0$\sigma^{-3}$). 
The systems are otherwise the same as in Fig.~\ref{fig:simul}(b).
The solid and dotted lines depict results based on the full APY theory and the superposition 
approximation, respectively. The density profile at a single solid-fluid interface ($L = \infty$) 
is also shown for comparison. 
\label{fig:n_z}
} 
\end{figure}

For comparison, we present in Fig.~\ref{fig:force2} the principal force components for a set of 
larger slit widths: $L= 3.45\sigma$, $3.60\sigma$, and $3.75\sigma$. There are no secondary 
maxima in this case. Instead we observe for 
$L= 3.60\sigma$ a broad region in the center of the slit where $F_{\uparrow}$ and 
$F_{\downarrow}$ virtually cancel each other and where, as a consequence, $dn/dz\approx 0$. 
Hence, this observation implies an essentially constant $n$ in the slit center, as can be seen in 
the third full curve of Fig.~\ref{fig:n_z}, where density profiles are shown for various cases.

The course of events shown in  Fig.~\ref{fig:force2} when we increase $L$ from $3.45\sigma$ 
to $3.75\sigma$ implies the formation of a layer at the slit center. The crossing of the principal force 
curves in Fig.~\ref{fig:force2}(a) at the slit center, $z_2=L/2$, corresponds to a density minimum, 
while that in Fig.~\ref{fig:force2}(c) corresponds to a density maximum. Note that for 
$L\approx 3.0\sigma$ 
there are four layers in the slit (two very sharp ones at the walls and two less sharp on either side of 
the slit center) and for $L\approx 4.0\sigma$ there are five layers. The fifth layer that forms in the 
middle for the intermediate separations arises via the broad flattening of the density profile in the 
middle, and signals the packing frustration in this case.

The data of Figs.~\ref{fig:force} and \ref{fig:force2} indicate a qualitative difference in $n(z)$ for 
$L \approx 1.65\sigma$ and $ \approx 3.60\sigma$. In the transition from $2 \rightarrow 3$ 
particle layers, the third layer is formed via the occurrence of secondary layers close to each 
surface, which merge to form a central layer with increasing $L$. 
This contrasts the transitions from $4 \rightarrow 5$ particle layers just discussed, where the 
new particle layer forms directly in the slit center. 
The secondary peaks for $L \approx 1.65\sigma$ are also evident in qualitatively different 
anisotropic structure factors $S({\bf q})$ for $L=1.60\sigma$ and $3.50\sigma$ presented in 
our previous work, Ref.~\onlinecite{nygard13}. 
$S({\bf q})$ for confined fluids is governed by an ensemble average of the anisotropic pair 
density correlations $n(z_1)h(z_1,z_2,R_{12})$ (see Ref.~\onlinecite{note_film} for more slit widths). 
In order to address these differences in $n(z)$ with $L$, we will in the following analyze 
further the principal force component $F_{\uparrow}$.

\subsection{Superposition approximation} 

\begin{figure}
\centering\includegraphics[width=8.5cm]{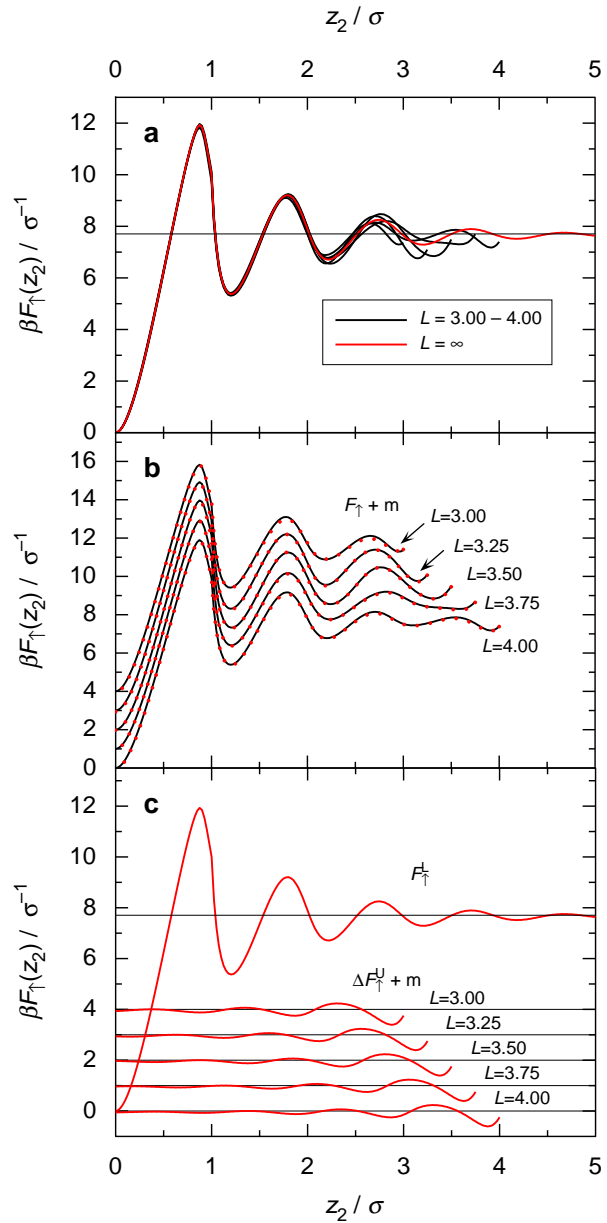}
\caption{Principal mean force component $F_{\uparrow}$ for the hard-sphere 
fluid between planar hard surfaces. The reduced separations are $L=3.00\sigma$, 
$3.25\sigma$, $3.50\sigma$, $3.75\sigma$, and $4.00\sigma$. 
The systems are otherwise the same as in Fig.~\ref{fig:simul}(b).
(a) $F_{\uparrow}$ for the confined fluids (black lines) and for a single 
solid-fluid interface ($L=\infty$, red line). 
(b) $F_{\uparrow}$ for the confined fluids (each offset vertically by $m=0... 4$ units for clarity), 
obtained via the full APY theory [solid lines, same as in (a)] and the superposition approximation 
(dotted lines). 
(c) Force components of the superposition approximation, $F_{\uparrow}^{\mathrm{L}}$ 
and $\Delta F_{\uparrow}^{\mathrm{U}}$, for different reduced slit widths (the latter curves are vertically 
offset by $m$ for clarity). $F_{\uparrow}^{\mathrm{L}}$ is the same as the red curve in (a).
\label{fig:F_up}
} 
\end{figure}

In both Figs.~\ref{fig:force} and \ref{fig:force2}, the principal force components $F_{\uparrow}$ 
(and $F_{\downarrow}$) for different $L$ nearly coincide for most $z$ values. In order to investigate 
this further, we plot $F_{\uparrow}$ for a wider set of slit widths, $L=3.0\sigma-4.0\sigma$, in 
Fig.~\ref{fig:F_up}(a). Indeed, apart from rather small deviations at large $z$, all data fall on a 
{\em master curve} given by $F_{\uparrow}$ for $L = \infty$, i.e., the force component for the 
single solid-fluid interface [the former curves are also shown separated in  Fig.~\ref{fig:F_up}(b)]. 
Although not shown here, we have verified that this observation holds reasonably 
well for $L \geq 1.0\sigma$, implying the same ordering mechanism irrespective of slit width.

In order to gain further insight into the ordering mechanism, we have determined density 
profiles obtained in a simple superposition 
approximation.\cite{percus80,snook81,wertheim90,sarman} 
Within this approximation, the potential of mean force $w$ in the slit is calculated as the sum of 
the corresponding potentials from two single hard surfaces, i.e. 
$w(z) \approx w_{\infty}(z) + w_{\infty}(L-z)$, where $w_{\infty}$ denotes the potential of mean 
force for the fluid at a single solid-fluid interface in contact with a bulk fluid of density 
$n_\mathrm{b}$. This implies  the superposition for the mean force: 
$F(z) \approx F_{\infty}(z) - F_{\infty}(L-z)$. Since the density profile is given by 
$n(z)=n_\mathrm{b}\exp[-\beta w(z)]$ the superposition approximation implies
\begin{equation}
n(z;L) \approx n^{\mathrm{sp}}(z;L)=\frac{n_{\infty}(z) n_{\infty}(L-z)}{n_\mathrm{b}},
\label{eq:n(z)sp} 
\end{equation}
where we have explicitly shown that the density profile for the slit, $n(z) \equiv n(z;L)$, 
depends on $L$, and where superscript sp indicates ``superposition'' and $n_{\infty}(z)$ is the 
density profile outside a single surface. 

In Fig.~\ref{fig:n_z} we compare $n(z)$ for reduced slit widths 
of $L=1.65\sigma$, $2.60\sigma$, and $3.60\sigma$ obtained via the full theory (solid lines) 
and the superposition approximation thus obtained (dotted lines). 
Note that there are density peaks at $z \approx 1.05\sigma$ for all three slit widths and that 
they  approximately coincide with the location of a density peak for the single solid-fluid interface 
(also shown in Fig.~\ref{fig:n_z}). This implies that the density peak at $z \approx 1.05\sigma$  
is strongly correlated with the bottom solid surface. 
Although the profiles obtained via the superposition approximation deviate quantitatively 
from those of the full theory, especially for narrow slit widths, the qualitative agreement implies 
that the main features of $n(z)$ -- the density peaks and shoulders of Fig.~\ref{fig:n_z} -- 
are rather uncomplicated confinement effects. 

To substantiate this conclusion, we present in Fig.~\ref{fig:F_up}(b) the principal force 
components $F_{\uparrow}$ for $L = 3.0\sigma - 4.0\sigma$, obtained both using the full theory 
and the superposition approximation. The agreement is equally good as for the density profile 
of the $L = 3.60\sigma$ case in Fig.~\ref{fig:n_z}. A significant point is now that the superposition 
allows us to separate the contributions to  $F_{\uparrow}$ from each surface in a simple manner, 
that will provide insights into what happens during confinement. As shown in 
Appendix~\ref{sec:appendixA}, $F_{\uparrow}$ can be decomposed in this approximation into 
two components: a major contribution from the lower surface, $F_{\uparrow}^\mathrm{L}$, and 
a correction due to the presence of the upper surface, $\Delta F_{\uparrow}^\mathrm{U}$. The former 
is the same as the average force component for the single solid-fluid interface plotted in 
Fig.~\ref{fig:F_up}(a) (denoted as ``master curve'' above). We have
\begin{equation}
F^{\mathrm{sp}}_{\uparrow}(z_2;L) = F_{\uparrow}^\mathrm{L}(z_2) + 
\Delta F_{\uparrow}^\mathrm{U}(z_2;L),
\label{eq:Fup_split} 
\end{equation}
where $\Delta F_{\uparrow}^\mathrm{U}(z_2;L)= F_{\uparrow}^\mathrm{U}(z_2;L) - 
F_{\uparrow}^\mathrm{b}$, see Eq.~(\ref{eq:Fup_splitA}). Here, $F_{\uparrow}^\mathrm{U}$ 
is the average force 
for the case of a single solid-fluid interface (U) and $F_{\uparrow}^\mathrm{b}$ is the force that acts 
on one side of a hard sphere (i.e. on one half) in the bulk fluid. Note that in 
$F^{\mathrm{sp}}_{\uparrow}$ it is only $F_{\uparrow}^\mathrm{U}$ that depends on $L$. 

\begin{figure}
\centering\includegraphics[width=6.5cm]{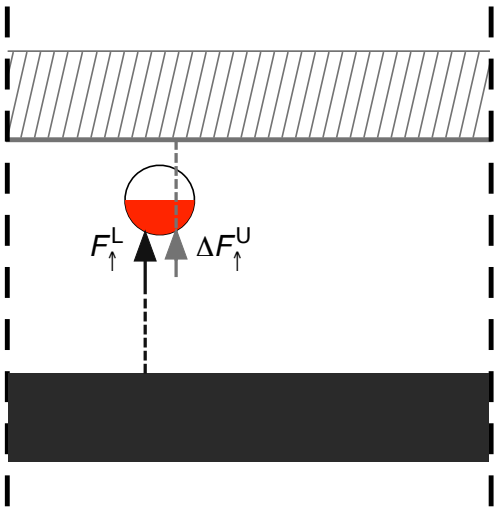}
\caption{A sketch illustrating the force contributions $F_{\uparrow}^\mathrm{L}$ and 
$\Delta F_{\uparrow}^\mathrm{U}$ in the superposition approximation. Each arrow represents 
a force that acts on the entire red half of the sphere (the location of the arrow has no significance 
in this sketch). The lower wall is shown in black and the upper wall is to be placed on the location 
indicated by the striped rectangle. The dashed line that connects each arrow to the respective 
surface indicates from which wall the influence originates. 
\label{fig:sketch_Fsp} 
} 
\end{figure}

In Fig.~\ref{fig:F_up}(c) we show $F_{\uparrow}^\mathrm{L}$ and $\Delta F_{\uparrow}^\mathrm{U}$ 
for the same surface separations as before. The $L$ dependence of the latter is simply a parallel 
displacement along $z$. When $F_{\uparrow}^\mathrm{L}$ and $\Delta F_{\uparrow}^\mathrm{U}$ 
are added we obtain the dotted curves in  Fig.~\ref{fig:F_up}(b). Thus the differences between 
each black curve and the red curve in  Fig.~\ref{fig:F_up}(a) is essentially contained in the 
contribution $\Delta F_{\uparrow}^\mathrm{U}$ from the upper surface (for smaller surface 
separations there 
will remain a minor difference as indicated by the small deviations for the superposition 
approximation in Fig.~\ref{fig:n_z}). 

To see in more detail what this means, we have in Fig.~\ref{fig:sketch_Fsp} shown schematically 
how these force contributions act on a sphere. In the presence of only one solid-fluid interface (L), 
the total force in the direction away from the surface (upwards) is 
$F_{\uparrow}=F_{\uparrow}^\mathrm{L}$, i.e., the force on the bottom half of the sphere shown 
as red in the figure. Let us now place the second surface (U) some distance from the other, at 
the location indicated in the figure. The change in the upwards force due to this second surface 
is given by $\Delta F_{\uparrow} \approx \Delta F_{\uparrow}^\mathrm{U}$ in the superposition 
approximation. Note that the former force, $F_{\uparrow}^\mathrm{L}$, acts on the hemisphere 
that is facing the surface L, while the latter, $\Delta F_{\uparrow}^\mathrm{U}$, is a force that acts 
on the hemisphere \emph{away from} the corresponding surface U and in the 
direction \emph{towards} this surface. 

If the lower wall were not present when we place the upper wall at the indicated position, the initial 
state would be a homogeneous bulk fluid and the final state a single solid-fluid interface (U) in 
contact with the bulk. In this situation $\Delta F_{\uparrow}^\mathrm{U}$ equals the actual change 
in the average force on the red hemisphere. In Eq.~(\ref{eq:Fup_split}) we have adopted this value 
as an approximation for the corresponding change when placing the upper wall in the presence of 
the lower one, i.e., when the initial state is an inhomogeneous fluid in contact with the lower 
surface and 
the final state is a fluid simultaneously affected by both surfaces. Since this approximation obviously 
is very good, it follows that the inhomogeneity due to one surface has only a small influence on 
the effects from the other surface throughout the entire slit. 

We saw in section~\ref{sec:Principal_comp} that the seemingly complicated changes in structure 
as the surface separation varies around half-integer $\sigma$ values of $L$ (i.e., $[m+0.5]\sigma$ 
with $m=$ integer), can be mainly 
explained by a parallel displacement of upward and downwards force curves along the $z$ direction. 
There was, however, some variation in these force curves near one of the surfaces (the upper 
surface for the upward forces and the lower surface for the downwards forces) that remained 
unexplained there. In the current section we have seen that this variation too can be mainly 
explained by a parallel displacement -- in this case a displacement of the contributions to 
$F_{\uparrow}$ (or $F_{\downarrow}$) due to each surface as seen in Fig.~\ref{fig:F_up}(c). 

To summarize our results in this section we make two important conclusions: 
First, by considering the mean force due to one surface (here the lower one) and by treating 
the influence from the other (upper) surface as a correction $\Delta F_{\uparrow}^{\mathrm{U}}$ 
according the the superposition approximation, one obtains nearly quantitative agreement with 
the full theory. Our 
approach of defining principal components of the mean force thereby provides a means to 
understand the contributions of each confining surface. 
Second, the principal force components obtained within the full theory and the superposition 
approximation are virtually in quantitative agreement for $L \geq 3.0\sigma$. For narrower slit widths 
(down to $L = 1.0\sigma$), quantitative discrepancies become more pronounced. These 
quantitative differences, which will be discussed in the next subsection, are nontrivial 
confinement effects. Nevertheless, the semi-quantitative agreement in the whole range of 
slit widths, down to $L = 1.0\sigma$, further strengthens the notion that ordering of confined 
hard-sphere fluids can, to a good approximation, be explained as a single-wall phenomenon. 
In essence, the fluid conforms locally with only one of the confining surfaces at a time. 
In some local regions it will thereby conform to one surface 
and in other regions to the other surface -- regions that are continuously changing (recall that the 
distributions we calculate are time averages of the various possible structures).
We emphasize 
that this reasoning holds for all slit widths, irrespective of whether $L$ is close to an integer or 
a half-integer multiple of the particle diameter (cf. Fig.~\ref{fig:F_up}). In other 
words, from a mechanistic point of view there is little difference between ordering in frustrated 
and more ordered confined hard-sphere fluids. In the latter case, the local ordering near one 
surface essentially agrees with the local ordering at the other one, whereby for the density 
profiles there appear only small mutual effects of the ordering from both surfaces beyond what is 
given by superposition. 

An interesting similarity between the structures observed in the fluid and solid phases should 
be mentioned. In some of the exotic crystalline structures observed under confinement 
-- most notably the prism-like structures~\cite{neser97,fortini06,oguz12} -- the particles locally conform with one of the solid surfaces. 
This is reminiscent of the situation in the fluid phase discussed above, 
although in the latter case the structures are less ordered and constantly changing locally.  
In particular, the adaptive prism phase $2P_A$ found in Ref. \onlinecite{oguz12} would yield an average density profile with secondary peaks on either side of the midplane, similar of those shown in Fig. \ref{fig:ng}(b) but much sharper.

The fact that the superposition approximation works surprisingly well for these rather large densities 
and gives a large part of the effects of confinement, means that it is simple to obtain good estimates 
of the density profiles for a confined fluid given an accurate density profile for a single solid-fluid 
interface. To obtain the latter is, however, computationally nontrivial and requires fairly advanced 
theories. Furthermore, as we shall see below, not all important properties of the confined fluid can 
be  explained by superposition.

\begin{figure}
\centering\includegraphics[width=8.0cm]{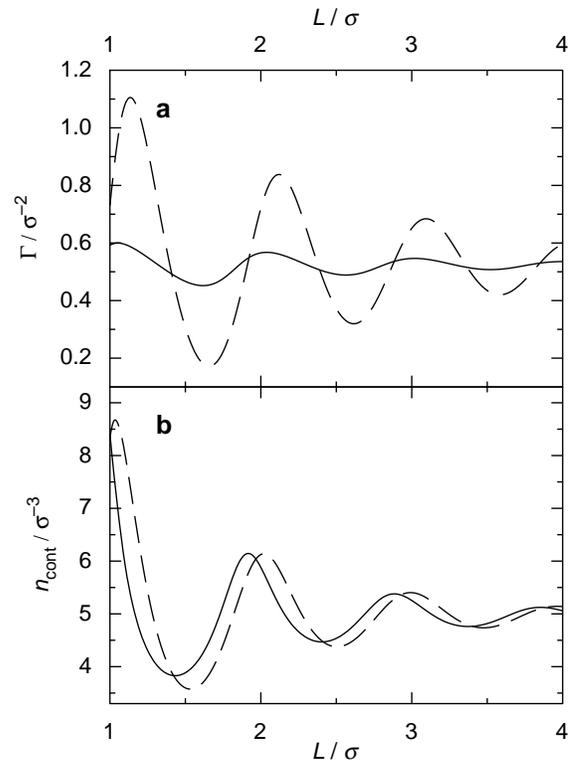}
\caption{(a) Excess adsorption $\Gamma$ and (b) contact density $n_{\mathrm{cont}} = n(0)$ of 
hard spheres between two hard planar surfaces as functions of reduced 
surface separation. The system is in equilibrium with a bulk fluid of density 
$n_{\mathrm{b}} = 0.75\sigma^{-3}$ [same as in Fig.~\ref{fig:simul}(b)]. 
Data are shown for both the full APY theory (solid lines) and the superposition 
approximation (dashed lines). 
\label{fig:contact} 
} 
\end{figure}

\subsection{Nontrivial confinement effects}

We have shown that the density profile $n(z)$ of confined hard-sphere fluids is, to a large 
extent, determined by packing constraints at a single solid-fluid interface. In this respect, the ordering 
is a trivial confinement effect. However, subtle deviations in $n(z)$ do remain in the superposition 
approximation, and these may lead to important, nontrivial confinement effects. The two most 
prominent nontrivial effects of confinement in, for example, Fig.~\ref{fig:n_z} are the slit width 
dependence of the contact density at the walls, $n_{\mathrm{cont}}$, and the total number of 
particles per unit area in the slit $N=\int_0^L n(z) dz$, which is a fundamental quantity for 
many properties of the confined fluid. In the following, we will discuss these two and related 
quantities in more detail. 

In Fig.~\ref{fig:contact}(a), we present the excess adsorption 
$\Gamma(L) = \int_0^L [n(z)-n_\mathrm{b}] dz$ of particles in the slit as a function of reduced slit 
width $L$, determined via both the full theory and the superposition approximation. 
The discrepancy between the two theoretical schemes is striking; since $\Gamma$ is an 
integrated quantity, minor systematic deviations in $n(z)$ accumulate to a large effect in the 
total number of particles. The superposition approximation gives, for example, in the interval 
$L=1.0\sigma$ to $2.0\sigma$ an estimate of $N$ that is wrong by a factor that varies between 
1.36 and 0.84.  
We note that, e.g., dynamic quantities such as diffusion coefficients\cite{mittal06} 
and relaxation times\cite{ingebrigtsen13} in simple confined fluids have been found to 
scale with particle packing, as quantified by the excess entropy. Consequently, a systematic 
error in the packing of particles (especially for very narrow confinement), as evidenced by 
systematic quantitative differences in the number density $n(z; L)$ and an ensuing large 
discrepancy in $\Gamma (L)$ between the full theory and the superposition approximation, will 
have a substantial impact on many properties of the confined fluid obtained theoretically.

Fig.~\ref{fig:contact}(b) shows the contact density 
$n_{\mathrm{cont}} = n(0)$ as a function of $L$, again obtained both via the full theory 
and the superposition approximation. This is an important quantity, because 
it yields the pressure between the walls, $P_{\mathrm{in}} = k_\mathrm{B}T n(0)$, according to 
the contact theorem. Consequently, $n_{\mathrm{cont}}$ is related to the net pressure acting on 
the confining surfaces, $\Pi(L) = P_{\mathrm{in}}(L) - P_{\mathrm{b}}$ with $P_{\mathrm{b}}$ 
denoting the bulk pressure, and hence to the extensively studied oscillatory surface 
forces.\cite{horn81,israelachvili} While the superposition approximation explains reasonably 
well the magnitude of $n_{\mathrm{cont}}$, there is a nontrivial systematic phase shift with 
respect to $L$ of about $0.1\sigma$. This effect has been observed by one of us (S.S.) already 
earlier,\cite{sarman} and in the following we will provide a mechanistic explanation of the 
phenomenon. A similar phase shift can also be seen in $\Gamma(L)$, Fig.~\ref{fig:contact}(a). 

In the superposition approximation, Eq.~(\ref{eq:n(z)sp}) yields the contact density for the 
wall at $z=0$ as 
\begin{equation}
n^{\mathrm{sp}}_{\mathrm{cont}}(L) =n^{\mathrm{sp}}(0;L)=\frac{n_{\infty}(0) 
n_{\infty}(L)}{n_\mathrm{b}}.
\label{eq:ncontact_sp} 
\end{equation}
Thus, the contact density for a reduced slit width $L$ is in this approximation proportional to the 
density at $z=L$ outside a single surface. To analyze the $L$ dependence further we will need 
the following equation that is equivalent to Eq.~(\ref{eq:LMBW}), 
\begin{multline}
\frac{d [\ln n(z_1)+\beta v(z_1)]}{dz_1}= \\ 
-\beta \int n(z_2)h(z_1,z_2,R_{12}) \frac{dv(z_2)}{dz_2} dz_2 d{\bf R}_{12}.  
\label{eq:LMBW-h}
\end{multline}
[The two equations can be transformed into each other by the Ornstein-Zernike 
equation~(\ref{eq:OZ}).]  For a single hard wall-fluid interface located at $z=0$, 
Eq.~({\ref{eq:LMBW-h}}) yields 
\begin{equation}
\frac{d n_{\infty}(z_1)}{dz_1} = n_{\infty}(z_1)n_{\infty}(0) \int h_{\infty}(z_1,0,R_{12}) d{\bf R}_{12}, 
\end{equation}
where $h_{\infty}$ is the total pair correlation function for the fluid outside the single surface.
By inserting $z_1=L$, this equation together with Eq.~(\ref{eq:ncontact_sp}) imply that 
\begin{equation}
\frac{d n^{\mathrm{sp}}_{\mathrm{cont}}(L)}{dL} =  n^{\mathrm{sp}}_{\mathrm{cont}}(L)
n_\infty(0) \int h_\infty(z_1,0,R_{12})   d{\bf R}_{12} \biggl \vert_{z_1=L}.   
\label{eq:contact_SP}
\end{equation} 
For the exact case, the corresponding equation can be obtained from Eq.~(\ref{eq:profile_equil}), 
which yields
\begin{equation}
\frac{d n_{\mathrm{cont}}(L)}{dL} =  [n_{\mathrm{cont}}(L)]^2 
 \int h(z_1,0,R_{12})   d{\bf R}_{12} \biggl \vert_{z_1=L}.  
  \label{eq:contact_exact}
\end{equation} 
Apart from the factors in front of the integral, we see that the main difference is that in the 
superposition approximation the total pair correlation function for a single wall is evaluated 
at coordinate $z_1=L$ outside the wall, while for the exact case the correlation function for the 
fluid in the slit is evaluated at the opposite surface (also at $z_1=L$). The oscillatory behavior 
of the contact density as a function of $L$ implies that its derivative changes sign with the 
same periodicity. Since the prefactors are positive, the phase shift for 
$n^{\mathrm{sp}}_{\mathrm{cont}}$ relative to $n_{\mathrm{cont}}$ must originate from the 
integrals.

\begin{figure*}
\centering\includegraphics[width=14cm]{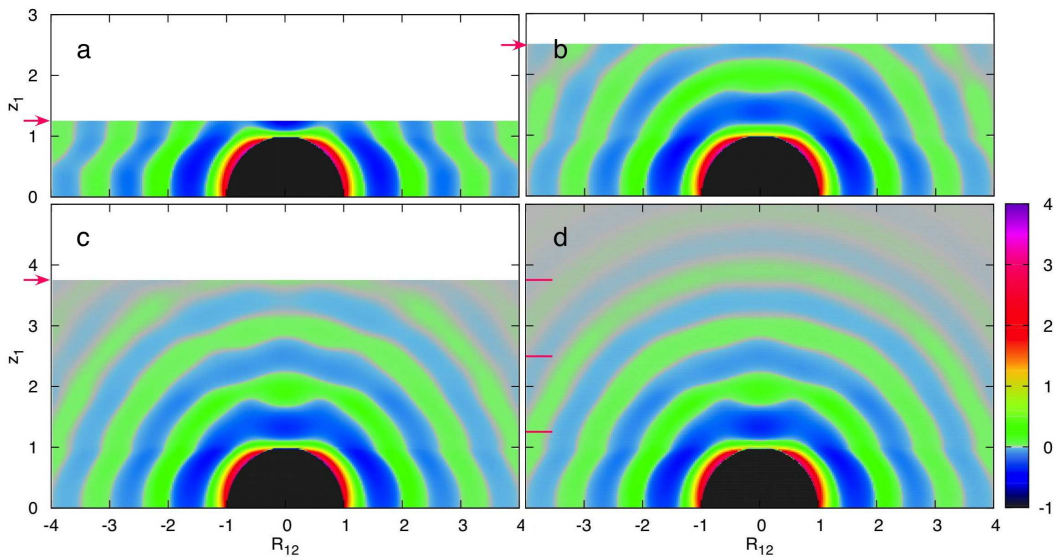}
\caption{Contour plot of the total pair correlation function $h(z_1,0,R_{12})$ at coordinate 
$({\bf R}_{12},z_1)$ around a particle located on the $z$ axis at coordinate 0, i.e., in contact with 
the bottom surface. 
Data are shown for different reduced slit widths: (a) $L = 1.25\sigma$, (b) $2.50\sigma$, 
(c) $3.75\sigma$, and (d) the single solid-fluid interface ($L=\infty$).
The systems are otherwise the same as in Fig.~\ref{fig:simul}(b).
 A small interval around $h=0$ is shown as gray in the contour scale and the black areas denote 
 the core region where $h=-1$. 
The red horizontal lines on the left hand side in (d) show the $z$ coordinate for spheres in contact 
with the top surface in subplots (a)$-$(c), i.e., at coordinate $z_1=L$ (cf. the red arrow in each of 
these subplots).  
\label{fig:h}
} 
\end{figure*}

In Fig.~\ref{fig:h} we have plotted the total pair correlation function $h(z_1,0,R_{12})$ in the 
slit when the central 
particle is in contact with the lower surface (i.e., at coordinate 0) for the cases $L = 1.25\sigma$, 
$2.50\sigma$, and $3.75\sigma$ together with the corresponding function for a single hard wall-fluid 
interface. The first impression is a striking similarity of these plots, despite that there is an 
upper surface 
present in the first three cases. There are only small differences in the entire slit compared to the single 
surface case for the corresponding $z_1$ values. When looking closely, one can, however, see some 
systematic differences in the $h$ function induced by the presence of the upper surface. Most 
importantly, we will investigate  $h$ for $z_1=L$, which occurs in the integral in 
Eq.~(\ref{eq:contact_exact}), and compare this with the values at the same $z_1$ coordinates for the 
single surface case, occurring in Eq.~(\ref{eq:contact_SP}). These  $z_1$ values are marked with red 
arrows in the left hand side of Figs.~\ref{fig:h}(a)--(c) and with red lines in Fig.~\ref{fig:h}(d).

\begin{figure}
\centering\includegraphics[width=8.0cm]{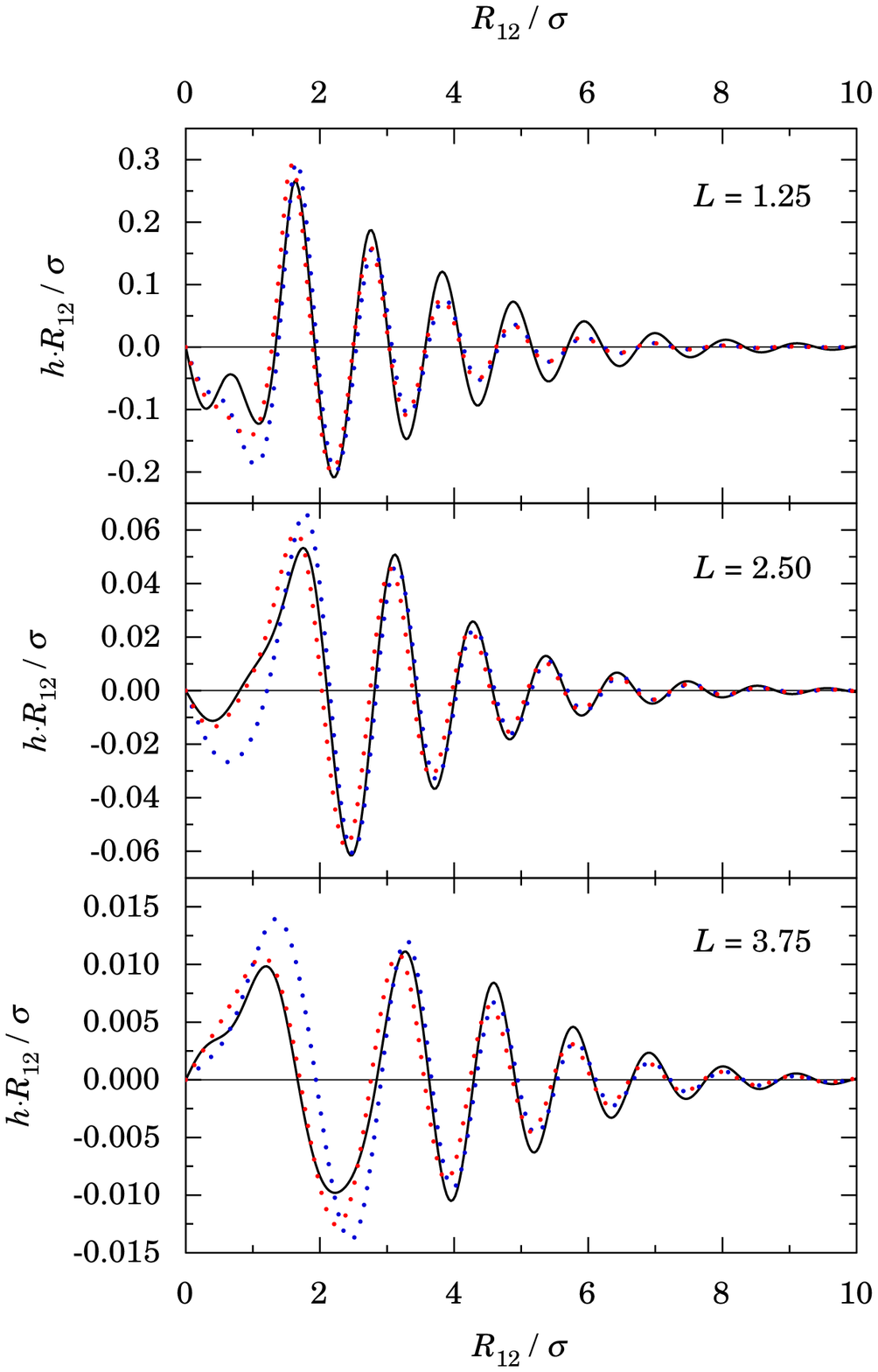}
\caption{$R_{12}\times h(L,0,R_{12})$ as function of $R_{12}$ for the systems in Fig.~\ref{fig:h}(a)-(c)  
with reduced surface separations 
$L=1.25\sigma$, $L=2.50\sigma$, and $L=3.75\sigma$. The data are obtained via full APY theory 
(solid line), superposition approximation (blue dotted line), and shifted superposition approximation 
($L \rightarrow L+0.1\sigma$, red dotted line). In the latter two cases, 
$R_{12}\times h_{\infty}(z_1,0,R_{12})$ is plotted for the appropriate $z_1$ values (see text). Note 
the different scales on the $y$ axis in the subplots. The curves go to zero at $R_{12}=0$ because  
of the factor $R_{12}$.
\label{fig:hR} 
} 
\end{figure}

Fig.~\ref{fig:hR} shows $R_{12}\times h(z_1,0,R_{12})$ with $z_1 = L$ for the cases in 
Figs.~\ref{fig:h}(a)--(c) and these curves are compared to $R_{12}\times h_{\infty}(z_1,0,R_{12})$ 
for the same $z_1$ coordinates (shown as blue dotted lines in the figure). The factor $R_{12}$ 
is  included so 
the areas under the curves in Fig.~\ref{fig:hR} are proportional to the values of 
the integrals of  Eqs.~(\ref{eq:contact_SP}) and  (\ref{eq:contact_exact}); 
this factor originates from the area differential $d{\bf R}_{12} = 2\pi R_{12} dR_{12}$. The $L$ values 
in Figs.~\ref{fig:h} and \ref{fig:hR} are selected such that we cover cases where 
$d n_{\mathrm{cont}}(L)/dL$ and $d n^{\mathrm{sp}}_{\mathrm{cont}}(L)/dL$ in 
Fig.~\ref{fig:contact}(b) are negative ($L = 1.25\sigma$) and positive ($L = 3.75\sigma$). There 
is also one case ($L = 2.50\sigma$) with $d n^{\mathrm{sp}}_{\mathrm{cont}}(L)/dL \approx 0$. 
These signs can be verified by inspection of the areas under the curves in Fig.~\ref{fig:hR} 
(the contributions around $R_{12}=0$ are most important for the sign; there are substantial 
cancellations in the tail region due to the oscillations). 

We can see in the figure that the full curves and the blue dotted ones do not agree, which means 
that the values of the integrals and hence of $d n_{\mathrm{cont}}(L)/dL$ are different, as expected. 
If we instead plot the values of $R_{12}\times h_{\infty}(z_1,0,R_{12})$ for  $z_1 = L+0.1\sigma$ 
(red dotted lines in the figure) we obtain better agreement. Thus the presence of the upper 
surface makes $h(z_1,0,R_{12})$ ``compressed'' in the $z$ direction by about $0.1\sigma$ 
compared to $h_{\infty}(z_1,0,R_{12})$. This compression gives rise to the phase shift observed 
in Fig.~\ref{fig:contact}. There are also some other small differences between  $h$ and $h_{\infty}$ 
and, in addition, there are different prefactors in Eqs.~(\ref{eq:contact_SP}) and 
(\ref{eq:contact_exact}). This gives the remaining differences in $n_{\mathrm{cont}}(L)$ 
and $n^{\mathrm{sp}}_{\mathrm{cont}}(L)$ seen in Fig.~\ref{fig:contact}(b).

The nontrivial confinement effects are accordingly due to rather delicate changes in the pair 
distribution function $g(z_1,z_2,R_{12})$ due to the presence of a second solid surface. The 
packing of particles in the slit around each individual particle is described by the pair density 
$n(z_1)g(z_1,z_2,R_{12})$ and the changes in 
$ng$ can be large, even for small variations in $g$, in regions where the density profile $n$ is 
large. Conversely, since there are large variations in the density profiles with surface separation, 
the packing is strongly altered even when the change in $g$ is small.

\section{\label{sec:summary}Summary and conclusions}

The self-consistent calculation of density profiles and anisotropic pair distribution functions, 
as provided by integral equation theories at the pair correlation level (like the APY theory used in 
this paper), gives efficient tools for the investigation of the structure of  inhomogeneous fluids 
in confinement. This is exemplified in this paper by a detailed examination of the mechanism 
behind the packing frustration for a dense hard-sphere fluid confined between planar hard walls 
at short separations. 

When the width of the slit between the walls is close to an integer multiple of sphere diameters, 
the layer structure is optimal and the number density profile $n(z)$ between the walls has sharp 
peaks. For slit widths near half-integer multiples of sphere diameters ($[m+0.5]\sigma$ with 
$m=$ integer),  the layer structure is much weaker and the packing frustration is large. The 
density profile shows considerable intricacy when the slit width is varied around these latter 
values. For example, when the reduced slit width $L$ is increased from $1.0\sigma$, there 
appear secondary density peaks close to the main peaks at each wall. These secondary peaks 
merge into a single peak at the slit center when $L$ approaches $2.0\sigma$. The mechanism 
behind these and other structural changes have been investigated in this paper, using the 
tools provided by the  anisotropic pair distribution function theory.

The number density profile is determined by the mean force $F(z)$ on the particles in the slit via 
the relationship $d\ln n(z)/dz =\beta F(z)$. For the hard-sphere fluid the mean force, which acts on 
a particle located at $z$, originates from collisions by other particles at the surface of 
the former. The average collisional force on the sphere periphery is proportional to the contact 
density there, which varies around the surface since the fluid is inhomogeneous. The sum of 
the average collisional forces constitutes the mean force $F$ and since we have access to the 
pair distribution, and thereby the contact density at the sphere surface, we can investigate the 
origin of any variations in $F$ and thereby in $n$. Of particular interest here are the variations 
when the slit width is changed. 

By introducing the two principal components $F_{\uparrow}$ and $F_{\downarrow}$ of $F$, 
each of which is the sum of the average collisional forces on the particle hemisphere facing one of the 
walls, we extract sufficient information from the pair distributions to obtain a lucid description of 
the causes for the structural changes due to varying degree of packing frustration. We show that 
most features of the structural changes, including the appearance and merging of the secondary 
peaks mentioned above, can be explained by a simple parallel displacement of the $F_{\uparrow}$ 
and $F_{\downarrow}$ curves when the slit width is varied around half-integer $\sigma$ values. 
The underlying reasons for this simple behavior is revealed via a detailed investigation of the 
pair distribution, that gives information about how the contact densities around the sphere 
periphery varies for different positions $z$ of a particle in the slit.  

It is found that the components $F_{\uparrow}$ and $F_{\downarrow}$, and thereby the ordering 
of the fluid, are essentially governed by the packing conditions at each single solid-fluid interface. 
The fluid in the slit  thereby conforms locally with only one of the confining surfaces at a time. 
In some local regions it will conform to one surface 
and in other regions to the other surface -- regions that are constantly changing (the 
calculated distributions are averages of the various possible structures).
This picture holds 
for all slit widths, irrespective of whether $L$ is close to an integer or 
a half-integer multiple of the particle diameter.

As a consequence of these local packing conditions, the force components $F_{\uparrow}$ 
and $F_{\downarrow}$, and thereby the total mean force $F=F_{\uparrow}-F_{\downarrow}$ acting 
on a particle in the slit, can to a surprisingly good approximation be written as a superposition 
of contributions due to the presence of each individual solid-fluid interface at the walls. When the 
slit width is varied, this superposition can be expressed in terms of a parallel displacement of 
force curves due to either surface. 

There are, however, some important properties of the inhomogeneous fluid that cannot be 
described by a simple superposition, but are instead determined by nontrivial confinement effects. 
In this paper, we exemplify such quantities by the number of particles per unit area in the slit $N$, 
the excess adsorption $\Gamma$, the contact density of the fluid at the wall surfaces $n(0)$,  and 
the net interaction pressure between the walls $\Pi$. In the superposition approximation, $N$ and 
$\Gamma$ disagree to a large extent compared to the accurate values, while $n(0)$ and $\Pi$ are mainly 
off by a phase shift in their oscillations. The analysis show that these nontrivial confinement effects 
are due to rather delicate changes in the anisotropic pair distribution function 
$g(z_1,z_2,R_{12})$ when the wall separation is changed.

\section*{Acknowledgments}
We thank Tom Truskett for providing the simulation data in Fig.~\ref{fig:simul}(a). 
K.N. and R.K. acknowledge support from the Swedish Research Council 
(Grant nos. 621-2012-3897 and 621-2009-2908, respectively). 
The computations were supported by the Swedish National Infrastructure for Computing 
(SNIC 001-09-152) via PDC.

\appendix
\section{\label{sec:appendixA}Force subdivision in superposition approximation}

For a hard sphere fluid in the slit between two hard walls, the force on, for example, the 
lower hemisphere of a hard sphere, $F_{\uparrow}$, can in  the superposition approximation 
be divided into contributions due to either wall surface. The contribution $F_{\uparrow}^\mathrm{L}$ 
from the lower surface is given by [cf. Eq. (\ref{eq:BGY})]
\begin{equation}
\beta F_{\uparrow}^\mathrm{L}(z_2)= 2\pi \int^{z_2}_{z_2-\sigma} dz_1 
n_{\infty}(z_1)g^{\mathrm{cont}}_{\infty}(z_1,z_2)(z_2-z_1),
\label{eq:BGY_L} 
\end{equation}
where $g^{\mathrm{cont}}_{\infty}$ is the contact value of the pair distribution for the fluid outside 
a single surface. Likewise, the contribution $F_{\uparrow}^\mathrm{U}$ from the upper surface is 
given by
\begin{multline}
\beta F_{\uparrow}^\mathrm{U}(z_2;L)= 2\pi \int^{z_2}_{z_2-\sigma} dz_1 
n_{\infty}(L-z_1) \\
\times g^{\mathrm{cont}}_{\infty}(L-z_1,L-z_2)(z_2-z_1).
\label{eq:BGY_U} 
\end{multline}
In the total $F^{\mathrm{sp}}_{\uparrow}$ there is a further contribution. From Eq.~(\ref{eq:n(z)sp}) 
we see that the total $\beta F^{\mathrm{sp}}$ is equal to the derivative of 
$\ln n^{\mathrm{sp}}(z;L) = \ln n_{\infty}(z) + \ln n_{\infty}(L-z) - \ln n_{\mathrm{b}}$. While the last 
term gives zero for $\beta F^{\mathrm{sp}}$, i.e., the mean force in bulk is zero, this is not the case 
for  $\beta F^{\mathrm{sp}}_{\uparrow}$. The mean force on one half of the sphere surface in 
bulk, $F_{\uparrow}^\mathrm{b}$, is non-zero; it is only the sum of the forces on both halves that 
are zero. Thus we have 
\begin{equation}
F^{\mathrm{sp}}_{\uparrow}(z_2;L) = F_{\uparrow}^\mathrm{L}(z_2) + 
F_{\uparrow}^\mathrm{U}(z_2;L) - F_{\uparrow}^\mathrm{b} 
\label{eq:Fup_splitA} 
\end{equation}
with $\beta F_{\uparrow}^\mathrm{b}= \pi \sigma^2 n_{\mathrm{b}} g^{\mathrm{cont}}_{\mathrm{b}}$, 
where $g^{\mathrm{cont}}_\mathrm{b}$ is the contact value for the pair distribution in bulk. When 
$L \rightarrow \infty$, the presence of the last term makes $F^{\mathrm{sp}}_{\uparrow}$ go to the 
single surface force $F_{\uparrow}^\mathrm{L}$, as it should in this limit.



%

\end{document}